\documentclass[%
 reprint,
 amsmath,amssymb,
 aps,
]{revtex4-1}

\usepackage{graphicx}
\usepackage{dcolumn}
\usepackage{bm}

\begin{document}

\preprint{APS/123-QED}

\title{Quantum transport with coupled cavities on the Apollonian network}

\author{Guilherme M. A. Almeida}

\author{Andr\'{e} M. C. Souza}%
 \email{Electronic address: amcsouza@ufs.br}
 
\affiliation{%
 Departamento de F\'{i}sica, Universidade Federal de Sergipe, 49100-000 S\~{a}o Crist\'{o}v\~{a}o, Brazil
}%

\date{\today}

\begin{abstract}
We study the dynamics of single photonic and atomic excitations in the Jaynes-Cummings-Hubbard (JCH) model
where 
the cavities are arranged in an Apollonian network (AN). 
The existence of a gapped field normal frequency spectrum along with
strongly localized eigenstates on the AN highlights many of the features provided by the model. 
By numerically diagonalizing the JCH Hamiltonian in the single excitation subspace,
we evaluate the time evolution of fully localized initial states, for many energy regimes.
We provide a detailed description of the photonic
quantum walk on the AN and
also address how an effective Jaynes-Cummings
interaction can be achieved at the strong hopping regime.
When the hopping rate and the atom-field coupling strength is of the same order, the excitation is relatively allowed to roam between atomic and photonic degrees of freedom as it propagates. However, 
different cavities
will contribute mostly to one of these components, depending on 
the detuning and initial conditions, in contrast to the strong atom-field coupling 
regime, where atomic and photonic modes propagate identically.
  
\begin{description}
\item[PACS numbers]
42.50.Pq, 42.50.Ex, 03.67.-a, 89.75.Da
\end{description}
\end{abstract}

\pacs{Valid PACS appear here}
\maketitle


\section{\label{sec1}Introduction}

Cavity quantum electrodynamics systems have been widely considered as a suitable choice
for implementation of quantum information processing schemes \cite{nielsen, pachos02, blais04, su08}.
The well-known Jaynes-Cummings (JC) model \cite{jaynes63} is the most important framework 
for investigating fundamental aspects of the interaction between light and matter.
It describes a two-state atom (qubit) coupled to a single field mode within a highly reflective cavity and it is exactly solvable, within the rotating-wave approximation. 
Further generalizations of this model include multiple atoms, $N$-level atom, dissipation, and so forth \cite{shore93}.

Experimental advances in optical microcavities \cite{armani03}, photonic crystals \cite{hennessy07}, 
and superconducting devices \cite{wallraff04} has brought interest to 
Hubbard-like models where photons can hop through an array 
of coupled cavity systems \cite{hartmann08rev, tomadin10}. An attractive
feature provided by such systems is the local addressing of individual cavities,
as long as the distance between adjacent cavities is larger than the optical wavelength of the resonant mode.  
In particular, 
the Jaynes-Cummings-Hubbard (JCH) model \cite{greentree06}  
provides a new framework for studying quantum many-body phenomena such as phase 
transitions where strongly correlated photons play the role.
Most of recent work has 
focused on the Mott insulator-superfluid quantum phase transition 
of polaritons (combined atomic and photonic excitations) \cite{greentree06, hartmann06, angelakis07, rossini07, aichhorn08, lehur09,
pippan09, schmidt09, knap10, hohenadler11}.

As coupled cavity systems allow the control and measurement of individual cavities, 
these turn out to be a reliable setup for distributed quantum information processing \cite{cirac99} as well, 
which requires entanglement generation \cite{angelakis07-ent} and quantum state transfer 
between distant nodes in a network \cite{cirac97, bose07-qst}. 
Some work has also addressed the propagation dynamics of single excitations
in such coupled cavity systems \cite{ogden08, makin09, ciccarello11, dong12}. 
These excitations can be photonic or atomic, where the first travels by direct hopping
and the latter by energy exchange with the photonic mode.  
By considering various energy regimes,    
Ogden \textit{et al}. \cite{ogden08} studied atomic state transfer
between two coupled cavities, each containing a single two-level atom.
They found that, by setting the appropriate detuning and hopping parameter,
a high-fidelity atomic excitation transfer from one cavity to another
can be achieved.
Makin \textit{et al}. \cite{makin09} investigated both atomic and photonic dynamics
in an one-dimensional array of coupled cavities. For limiting 
energy regimes, they showed that the system can be mapped to two uncoupled
Heisenberg spin chains. The dynamics turns to be much more complex when the hopping rate
and atom-field coupling parameter is of the same order. 
They have also shown that the pulse dispersion can be 
attenuated by considering a parabolic distribution of hopping rates, 
or by encoding the initial state as a Gaussian wave packet.
In \cite{dong12}, Dong \textit{et al}. discussed the binary transmission dynamics 
in a chain of coupled cavities, each containing a tree-level atom, and 
provided a class of encoding protocol which can improve the state transmission fidelity.
In \cite{ciccarello11}, Ciccarello addressed the appearance of an effective JC interaction at the strong hopping regime, for large-size arrays.
By considering a staggered pattern of hopping rates, 
instead of uniform, a gapped discrete field normal frequency in the
middle of the band is induced, corresponding to a strongly localized
field normal mode. By setting the appropriate parameters,
a significant atom-field interaction can persist even 
if we increase the system size and thus the atomic excitation is no longer frozen.

These results indicate that even in the single excitation subspace, 
the JCH model presents an appealing dynamics due to the interplay between atomic 
and photonic degrees of freedom. So far, only regular
structures have been considered. Thus it raises the question of whether new
features arise if we arrange the cavities in more complex structures.

\begin{figure}[t] 
\includegraphics[width=0.4\textwidth]{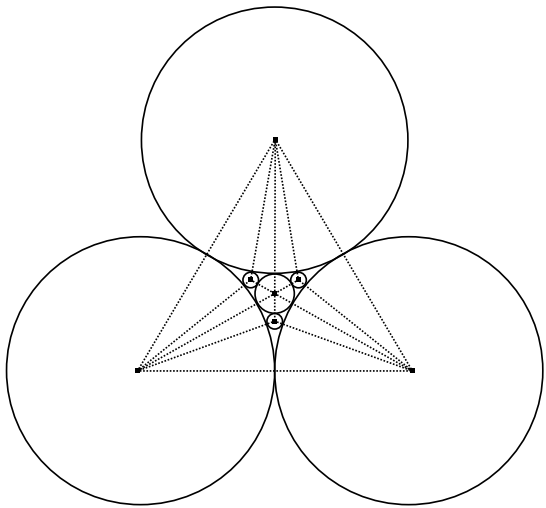}
\caption{\label{fig1} The second generation ($n=2$) Apollonian packing (solid lines)
and the corresponding AN (dotted lines).}
\end{figure}

Lately, a class of networks which are neither completely regular nor
completely random, named small-world networks \cite{watts98},
where the average length of the shortest path $\textit{l}$ between 
two nodes increases logarithmically with the network size $N$ ($\textit{l}\varpropto \mathrm{ln}N$),
has been widely studied within many fields, from social and biological systems \cite{watts98, newman01}
to classical and quantum transport dynamics \cite{latora01, kim03, mulken07}. 
In particular, a class of complex networks called Apollonian networks (AN) \cite{andrade05}, 
which are simultaneously small-world, scale-free (that displays a power law degree distribution), 
and can be embedded in Euclidean space, has brought some attention. 
The free-electron gas \cite{oliveira09}, magnetic models \cite{andrade05mag}, 
tight-biding models \cite{cardoso08}, correlated electron systems \cite{andrem07},
and quantum walks \cite{xu08} in such networks have been investigated. 
AN's arise from the problem of space-filling packing of spheres \cite{boyd73}. 
It can be deterministic or random, and depends on the initial configuration. 
The most common packing starts with three mutually touching circles.
The first generation, $n=1$, is achieved by inserting a maximal circle 
in the interstice bounded by the three initial circles. 
Further generations are obtained by repeating this process for every empty space.
This packing can be mapped to the AN as shown in Fig. \ref{fig1}.
At a given generation, the number of nodes is given by $N(n)=(3^{n}+5)/2$, 
and the number of connections by $C(n)=(3^{n+1}+3)/2$ ($n=0,1,2,...$). In each generation a new group of nodes sharing 
the same degree $\gamma$ is created. More precisely, there are $3^{n-1}, 3^{n-2}, 3^{n-3},..., 3^{2}, 3, 1$, and $3$ nodes with 
degree $\gamma =3,3\cdot 2, 3\cdot 2^{2}, ..., 3\cdot 2^{n-2}, 3\cdot 2^{n-1}$, and $2^{n}+1$, respectively, where 
the last number represents the three corner nodes. The central node,
often referred as the \textit{hub node}, has the largest degree.    

In this paper, we study the dynamics of the JCH model where the cavities are
arranged in such AN. 
We numerically diagonalize the Hamiltonian in the single excitation subspace
and solve the Schr\"{o}dinger equation
thus obtaining the time evolution of 
atomic and photonic occupation probabilities in every cavity,
for fully localized initial states with one excitation only.
We consider three different energy regimes. For the strong hopping case, the
photonic component propagates practically free, without significantly exchange energy with the 
atomic modes in short time scales. Also, a JC-like interaction
can be triggered by setting the appropriate 
atomic transition frequency and initial conditions.
For comparable hopping and atom-field coupling, 
both JC-like dynamics and/or photonic quantum walk 
on the AN may be disturbed by the appearance of
more complex JCH eigenstates (polaritons). 
Thus, each cavity might be able to sustain mostly one of
both components.  
When the atom-field coupling strength
dominates, atomic and photonic dynamics become identical.  

In what follows, Sec. \ref{sec2}, we introduce the JCH model. In Sec. \ref{sec3} we make an analysis of the free-field Hamiltonian spectrum and eigenstates. 
In Sec. \ref{sec4} we evaluate the time evolution for a variety of energy regimes and discuss the overall dynamics
in terms of the JCH eigenstates. 
Finally, in Sec. \ref{sec5}, we draw
our conclusions. 

\section{\label{sec2}The Jaynes-Cummings-Hubbard model}

Let us consider a system of coupled cavities where each occupies the nodes of an Apollonian network.
A two-level system $\lbrace \left| g \right>, \left| e \right> \rbrace$ coupled 
to a single field mode can be described by the JC Hamiltonian \cite{jaynes63, shore93}
(in the rotating wave approximation),
\begin{equation}
h_{i}^{JC}=\dfrac{\omega_{a}}{2}\sigma_{z}+\omega_{f}a_{i}^{\dagger}a_{i}+\beta 
(\sigma_{i}^{+}a_{i}+\sigma_{i}^{-}a_{i}^{\dagger}),
\end{equation}
where $\sigma_{z} \left| g \right> = - \left| g \right>$ and
$\sigma_{z} \left| e \right> =  \left| e \right>$, 
$\sigma_{i}^{+}$ ($ \sigma_{i}^{-} $) and $ a_{i}^{\dagger} $ ($a_{i}$) 
are, respectively, the atomic and photonic raising (lowering)
operators for cavity $i$, $\beta$ is the atom-field coupling strength, 
$\omega_{a}$ the atomic transition frequency, $\omega_{f}$ the field mode frequency,
 and we set $\hbar = 1$ for convenience. 
If we include photon evanescent hopping between nearest-neighbour cavities 
(tight-binding approximation), we can represent the whole system by the JCH Hamiltonian  
\begin{equation} \label{jchhamilt}
H=\sum_{i=1}^{N(n)}h_{i}^{JC}- \kappa\sum_{i,j=1}^{N(n)} A_{ij}^{(n)}a_{i}^{\dagger}a_{j},
\end{equation}
where $\kappa$ is the inter-cavity hopping rate, 
$N(n)$ number of nodes for a given AN generation $n$, 
and $A_{ij}^{(n)}$ the adjacency matrix elements. We now restrict the system basis to the 
single excitation subspace. These excitations
can be photonic or atomic, that is $ \left| k \right> \left| g, 1 \right>$ or $ \left| k \right> \left| e, 0 \right>$, respectively, where 
$k\in \left\lbrace 1,...,N(n) \right\rbrace $ specifies the cavity. As such, the full system state 
is written as a tensor product of every
single cavity state, where one excitation must be conserved (the other non-excited cavities are 
in the $\left| g,0 \right>$ state), so the Hilbert space has $D=2N(n)=3^{n}+5$ dimensions. 
In such basis, the matrix form of the Hamiltonian (\ref{jchhamilt}) can be simply 
expressed by \cite{makin09}
\begin{equation} \label{hamilt1exc}
H_{1exc} =  \dfrac{\Delta}{2} I_{N(n)} \otimes Z + \beta I_{N(n)}
\otimes X -\kappa A_{n} \otimes 
\dfrac{I_{2}+Z}{2},
\end{equation} 
where $\Delta = \omega_{f} - \omega_{a}$ is the detuning between photonic and atomic
modes, $I_{m}$ is the $m \times m$ identity matrix, $X$ and $Z$ are the usual Pauli matrices, 
and $A_{n}$ the adjacency matrix.
The ground state $\otimes_{k=1}^{N(n)}\left| k \right> \left| g, 0 \right> $ is not  
included in this subspace, that is, the system is already set with
a single excitation. 

\begin{figure}[t] 
\includegraphics[width=0.48\textwidth]{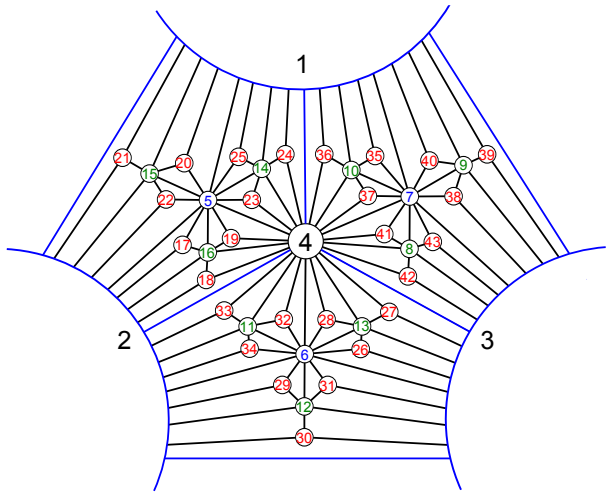}
\caption{\label{fig2} (Color online) The Apollonian network for $n=4$ (43 nodes). The first generation is made up by nodes 1 to 4, the second one includes 5 to 7, the third, 8 to 16, and the fourth, 17 to 43. The corner nodes were drawn bigger for a clearer visualization. This is the node labelling we will adopt through the paper.}
\end{figure}

The AN does not take geometrical 
deformations into account so we are only interested in the connections between the nodes.
By considering uniform inter-cavity coupling, the adjacency matrix $A_{n}$ for $n=2$,
for example, can be written by 
\begin{equation}
A_{2} = \begin{pmatrix} 0 & 1 & 1 & 1 & 1 & 0 & 1  \\ 1 & 0 & 1 & 1 & 0 & 1 & 1 \\ 1 & 1 & 0 & 1 & 0 & 0 & 0 \\ 1 & 1 & 1 & 0 & 1 & 1 & 1 \\ 1 & 0 & 0 & 1 & 0 & 0 & 1 \\ 0 & 1 & 0 & 1 & 0 & 0 & 1 \\ 1 & 1 & 0 & 1 & 1 & 1 & 0 \end{pmatrix}
\end{equation}
(see Fig. \ref{fig1}). 
In order to discuss our results in the following sections, we need the appropriate identification 
of each node (cavity), then Fig. \ref{fig2} provides the node numbering we are going 
to consider herein, up to the 4th generation AN.
%

\section{\label{sec3}Spectrum and localization properties}

The JCH model eigenstates are known as polaritons, superposition of atomic and photonic excitations over the entire lattice. As we are considering
the single excitation subspace, the polaritons are linear combinations written on the 
$\lbrace \left| k \right> \left| g, 1 \right>, \left| k \right> \left| e, 0 \right>\rbrace$
basis, as seen in the last section. 

\begin{figure}[t]
\includegraphics[width=0.40\textwidth]{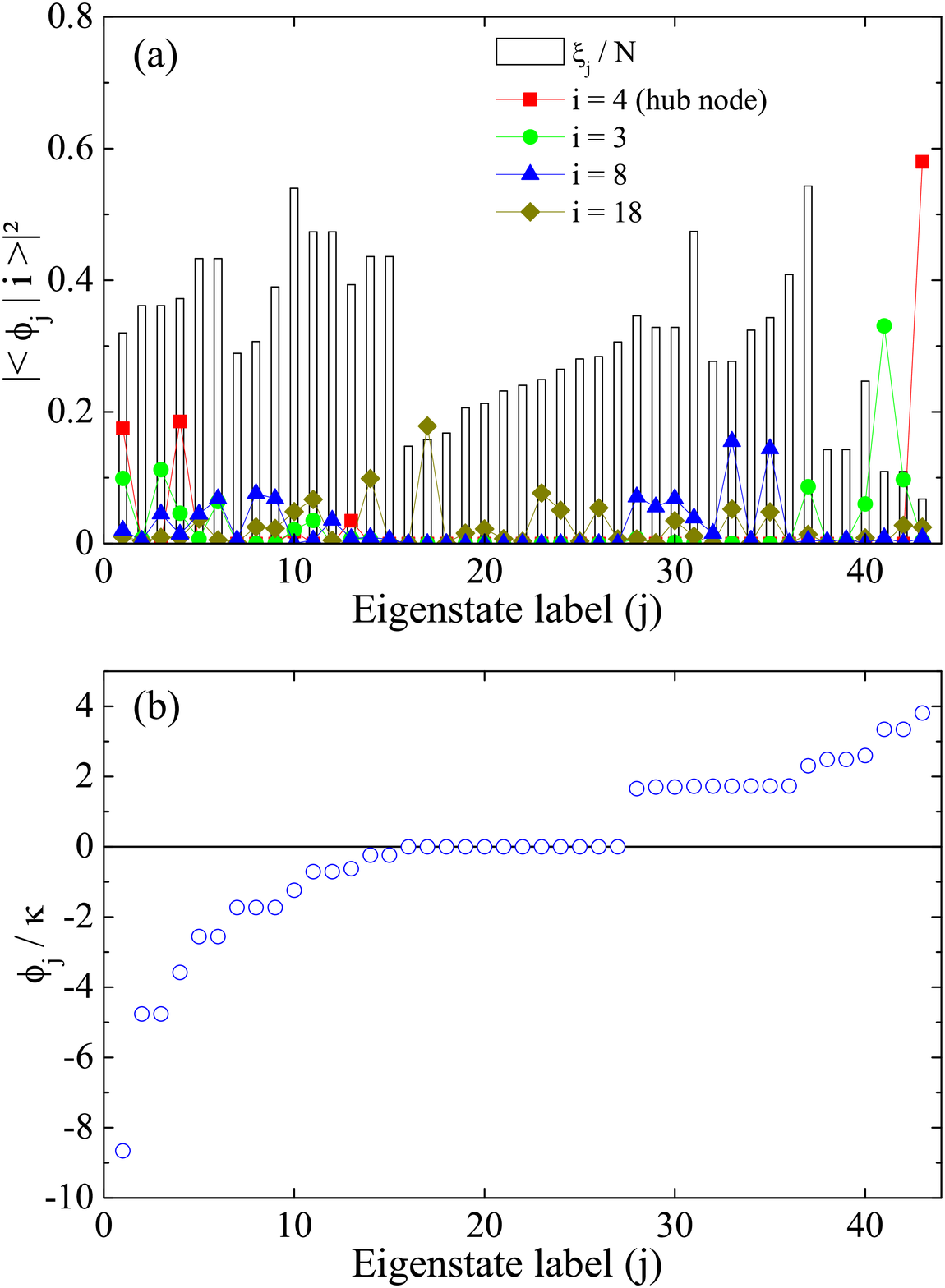}
\caption{\label{fig3} (Color online) (a) The normalized participation ratio of eigenstates $\xi_{j}/N$ (column bars) of
$H_{field}$ and 
the node coefficient amplitudes, 
$\left| \left< \phi_{j} \vert i \right> \right| ^{2}$, for $i = 4$, 3, 8, and 18
on the $n=4$ AN (43 nodes).  
(b) The field normal mode frequency band ($\omega_{f} = 0$).
The eigenstates are labelled according to increasing values of $\phi_{j}$
(in units of $\kappa$)
and, within each degenerate group, by increasing values of $\xi_{j}/N$.} 
\end{figure}

The AN topology induces both localized and extended eigenstates, and its spectrum
is characterized by several regions, some of these with a high degree of degeneracy due to  
the node degree distribution (see Fig. \ref{fig2}). Such configuration drastically changes the  
the way atomic and photonic degrees of freedom relate with each other for specific sets of parameters, 
comparing to the JCH model defined in regular lattices. 
Hence, in order to get an idea of the form such polaritons will take,
we shall provide a brief description of 
the free-field Hamiltonian,   
\begin{equation}
H_{field}= \omega_{f}\sum _{i=1}^{N(n)} a_{i}^{\dagger}a_{i}  -\kappa \sum_{i,j=1}^{N(n)} A_{ij}^{(n)}a_{i}^{\dagger}a_{j},
\end{equation}
normal frequency spectrum and eigenstates on the single photon basis. We solve the above Hamiltonian by numerical diagonalization. 
In order to characterize the degree of localization of such eigenstates we evaluate the participation ratio, defined 
as
\begin{equation}
\xi _{j}=\dfrac{1}{ \sum\limits_{i=1}^{N(n)}
\left| \left< i \vert \phi_{j} \right> \right| ^{4}},
\end{equation}
for a given AN generation $n$, where $\vert \phi_{j} \rangle$ represents an eigenstate and $\vert i \rangle$ a single photon located at node $i$.
This quantity can assume values within $1$, for completely localized states, and $N$, for
extended states ($\left< i \vert \phi_{j} \right> = 1/\sqrt{N}$ for all $i$). In Fig. \ref{fig3}(a) 
we compare the participation ratio with the 
node coefficient amplitudes,
$\left| \left< \phi_{j} \vert i \right> \right| ^{2}$, for different $i$'s
on the 4th generation AN. It gives us an outlook on the way these states are distributed along the spectrum.  
The associated field normal mode frequencies $\phi_{j}$ are shown in Fig. \ref{fig3}(b), for $\omega_{f}=0$ (in units of $\kappa$).     
%
Firstly, notice that the hub node $i=4$, with degree $\gamma = 24$, has its largest amplitude 
in the most localized eigenstate, $\xi_{j = 43}/ N \approx 0.07$, corresponding to the   frequency $\phi_{j=43}/ \kappa \approx 3.8$,
which is non-degenerate. From now on, we call it $\phi^{hub}$.   
For the other nodes, $i=3$, 8, and 18, representing the $\gamma = 17$, 6, and 3 group,
respectively, more 
eigenstates get involved but there is still a peak over a reasonable localized eigenstate.
It turns out that a similar distribution occurs for every node $i$, with different
localization strengths, not necessarily scaling with $\gamma$, although
the hub node localized eigenstate will always be paramount, regardless of the network size. 
Moreover, even within a node group 
sharing the same $\gamma$, there can be a difference in such localization
strengths,
since there can be many symmetry groups in it (the AN has a $2\pi /3$ rotational symmetry).  
Anyway, what is most relevant to be addressed here is that,
apart from the hub node localized eigenstate that corresponds 
to a non-degenerate frequency $\phi^{hub}$, 
every node has a (not solely) related localized eigenstate associated 
with a degenerate frequency level ($\phi_{j} / \kappa \geq 0$) depending on $\gamma$ 
(higher frequencies are associated with higher $\gamma$ values). By looking at Figs.
\ref{fig3}(a) and \ref{fig3}(b)
we can identify the level
$\phi_{j} / \kappa = 0$ as being related to the
$\gamma = 3$ group (nodes 17 to 43),  
for instance. Frequencies below this value comprise extended eigenstates mostly. 
   
Although we have only discussed the $n=4$ AN spectrum properties, 
the AN structure presents self-similarity as the generation increases and thus it 
reflects on its properties. 
These spectrum regions containing strongly localized eigenstates for a given $\gamma$
yield to attractive features when we consider the full JCH Hamiltonian as
we are going to show in the next section. 
Further details about localized and extended states on the AN and its relationship with 
the structure size and characteristic spectrum, for a similar tight-binding Hamiltonian, 
can be found in Ref. \cite{cardoso08}.

\section{\label{sec4}Time evolution}
 
In this section we display the results for time evolution of the JCH model 
arranged on the considered AN. The system eigenstates $\left| E_{j} \right>$ (polaritons)
and its corresponding eigenvalues $E_{j}$ were evaluated
by means of numerical diagonalization
of Hamiltonian (\ref{hamilt1exc}). For an initial state of the form  
$\left| \psi _{k}(0) \right> = \left| k \right>\otimes (\cos\alpha \left| g, 1 \right> + 
\sin\alpha \left| e, 0 \right>)$ fully localized at node (cavity) $k$, we discuss the system
dynamics for different energy regimes.
The state at time $t$ is given by
$\left| \psi(t) \right>=U(t)\left| \psi(0) \right>$, where $U(t)=e^{-iHt}$ is the quantum time evolution
operator. 
We are primarily interested on the single excitation propagation dynamics. Then for a given initial state
$\left| \psi _{k}(0) \right>$, we write
\begin{equation}
\pi _{\ell k}^{ph}(t)=\left| \sum_{j}^{D} e^{-iE_{j}t}
\left< \psi _{k} (0) \vert E_{j} \right>\left< E_{j} \vert \ell, g \right> \right| ^{2},
\end{equation}
\begin{equation}
\pi _{\ell k}^{at}(t)=\left| \sum_{j}^{D} e^{-iE_{j}t}
\left< \psi _{k} (0) \vert E_{j} \right>\left< E_{j} \vert \ell, e \right> \right| ^{2},
\end{equation}
as the probability of finding the photonic and atomic excitations, respectively, 
at node $\ell$ in time $t$, where we have denoted 
$\left| \ell, g \right> \equiv \left| \ell \right>\left| g, 1 \right>$  and 
$\left| \ell, e \right> \equiv \left| \ell \right>\left| e, 0 \right>$.

\subsection{Strong hopping regime}
   
Firstly, we investigate the strong hopping regime, $\kappa \gg \beta$. In this case,
the existence of discrete field normal mode frequencies along with
strongly localized eigenstates 
brings an interesting feature: a JC-like
interaction between the field normal mode and 
its atomic analog occur 
for an appropriate tuning of the atomic transition frequency, $\omega_{a}$, 
regardless of the system size \cite{ciccarello11}, since the AN maintain 
its spectrum pattern as the 
generation increases. 
It means that the atomic component still can propagate
(in a time scale of the order of $\beta$) even in such a regime where the atom-field interaction rate is much slower than the
photon inter-cavity hopping. 
This property comes from the fact that, as long as
the single cavity parameters are uniform through the lattice and the decoupling of the 
hopping Hamiltonian in terms of normal modes is known,
the JCH Hamiltonian can be rearranged into a sum of decoupled JC models, each
coupling a field normal mode to its atomic counterpart \cite{ogden08, makin09, ciccarello11}.
%
\begin{figure}[t]
\includegraphics[width=0.48\textwidth]{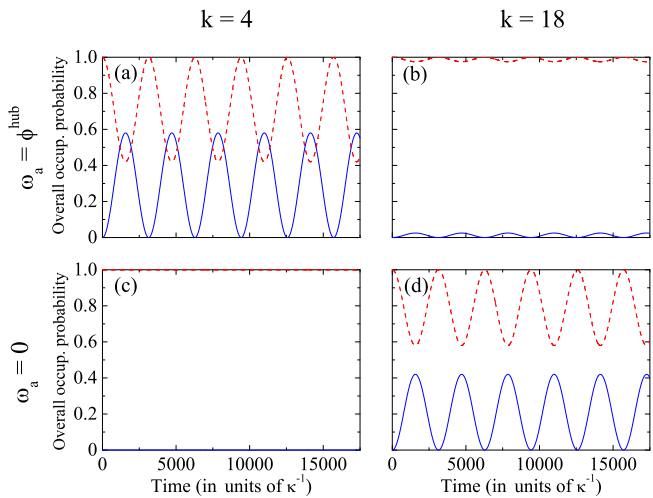}
\caption{\label{fig4} (Color online) Photonic (solid line) and atomic (dashed line) overall occupation probabilities,
$\sum_{\ell} \pi_{\ell k}^{type}$,  
for a fully atomic initial state $\left| \psi _{k}(0) \right> = \left| k, e \right> $ on the $n=4$ AN.
System parameters are $\kappa / \beta =10^{3}$ and $\omega_{f} = 0$.}  
\end{figure}
%
%
\begin{figure}[t]
\includegraphics[width=0.45\textwidth]{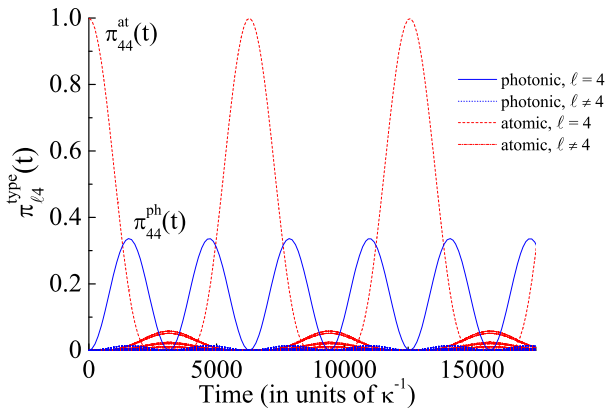}
\caption{\label{fig5} (Color online) Photonic and atomic occupation probabilities for a fully atomic initial state $\left| \psi _{4}(0) \right> = \left| 4, e \right> $ on the $n=4$ AN. Most relevant probabilities are indicated on the figure.
System parameters are $\kappa / \beta =10^{3}$, 
$\omega_{a}=\phi^{hub}$, and $\omega_{f} = 0$.
} 
\end{figure} 
In other words, when $\kappa \gg \beta$, 
we have well defined atomic- and photonic-like polaritons, i.e., eigenstates
where
$\left<  i, g \vert E_{j} \right>$ and $\left<  i, e \vert E_{j} \right>$ 
is null, respectively, for every node $i$. 
By a judicious tuning of the atomic frequency to a discrete non-denegerate field normal mode frequency, 
a pair of such polaritons 
starts to overlap with both components, by a similar way as the single cavity JC model eigenstates \cite{jaynes63, shore93} except that,
in the JCH model, it extends over the entire network. 
If the field normal mode frequency happens to be $n$-fold degenerate, then 
$n$ pairs of polaritons get involved in such process. Although in this case there is no JC-like interaction
exclusively between one field normal mode and its atomic analog, a significant energy transfer between
atomic and photonic degrees of freedom might still occur as long as 
a pair of localized JC-like polaritons
is available and we set the appropriate initial state.
In order to highlight these phenomena, mainly to address the amount of energy that can be exchanged between atomic
and photonic components, for given initial conditions,
we shall start the system with a fully localized atomic state.  
As such, we induce the system dynamics to be generated only by those
JC-like polaritons, for the given atomic frequency, and 
thus the photonic component does not propagate freely. 
In Figs. \ref{fig4}(a) and \ref{fig4}(b) we show the overall photonic (atomic) occupation probability, 
$\sum_{\ell} \pi_{\ell k}^{ph}$ ($\sum_{\ell} \pi_{\ell k}^{at}$), 
when the system starts with an atomic excitation at nodes 4 (hub node) and 18, respectively.
The atomic frequency $\omega_{a}$ was set to $\phi^{hub}$ in order
to simulate the dynamics of 
the JC-like polaritons strongly localized at the hub node.
Observe that a significant atom-field interaction happens
only when the atomic excitation is initially present at such node. In Fig. \ref{fig5}
we show how the process takes place in a detailed way.
The initial concentrated atomic energy is progressively converted into field modes (mostly
at node 4) and it reaches the other cavities as well. Right after the field modes reach its maximum, it releases energy until the point that the system is fully atomic
again,
but the excitation is delocalized instead of being fully localized at the initial node. Then a reverse process occur, i.e., the field modes retrieve the
same amount of energy and transfer all of it to the node 4 in the form of atomic excitation, 
thus recovering the initial state.
As long as $\kappa / \beta$ is high enough and $\omega_{a}$ is precisely adjusted to the discrete 
field normal mode frequency, the same behavior occurs cyclically. 
In Figs. \ref{fig4}(c) and \ref{fig4}(d), the atomic frequency matches
with the degenerate field normal mode frequency level comprising every 
localized polariton 
associated to the $\gamma=3$ node group. 
In this case, as we prepare the atomic excitation at node 4,
it freezes [Fig. \ref{fig4}(c)], indicating that the overlap between 
this node and every involved JC-like polaritons is minimum. The same does not applies
to node 18 and thus a considerable atom-photon interaction is triggered [Fig. \ref{fig4}(d)]. 
Even though now there are many polaritons taking part on the dynamics (as the discrete
field normal mode frequency level is degenerated), the atom-field interaction occurs in a similar way
as in Fig. \ref{fig5}, since most of the contribution comes from the localized polaritons.  
Thus, despite of the degeneracy degree of a discrete field normal mode frequency level,
we can assert that the 
overlap between the involved JC-like polaritons and the initial
state 
dictates the amount of energy in which can be exchanged.
This property is still valid for large-size networks 
since the AN field normal frequency band is characteristically gapped for any generation.
Observe that, in principle, such effective JC interaction is possible in small regular clusters with uniform hopping rates, for instance. 
However, as we increase
its size, the field normal spectrum tends to a continuum thus making an accurate resonance not achievable anymore.
In addition, the involved polaritons 
become even more extended. Therefore, 
it does not matter in which node we set the initial state, 
the atomic component becomes practically frozen \cite{makin09}.    

Let us now consider an initial state of the form 
\begin{equation} \label{scstate}
\left| \psi _{k}(0) \right> = \left| k \right>\otimes (\left| g, 1 \right> + 
\left| e, 0 \right>)/\sqrt{2},
\end{equation}
which is a JC model single cavity eigenstate \cite{jaynes63, shore93}.
In this case, the whole frequency band is available 
for the photonic mode propagation while the atomic mode behavior depends
on the atomic frequency $\omega_{a}$, in a similar way as in 
the previous discussion where the initial state was fully atomic. Thus,
the photonic mode describes a quantum walk through the network \cite{xu08} while
the atom-field interaction takes place in a much longer time scale,
since $\kappa \gg \beta$. Obviously, the latter is only possible if 
the atomic frequency matches with any of the field normal mode frequencies.  
Anyway, regardless of $\omega_{a}$, in this case the photonic
mode propagates practically free without being disturbed by the atom-field
interaction. Hence, in the remaining of this section we discuss
the quantum transport properties of the photonic excitation on the AN.

\begin{figure}[t]
\includegraphics[width=0.47\textwidth]{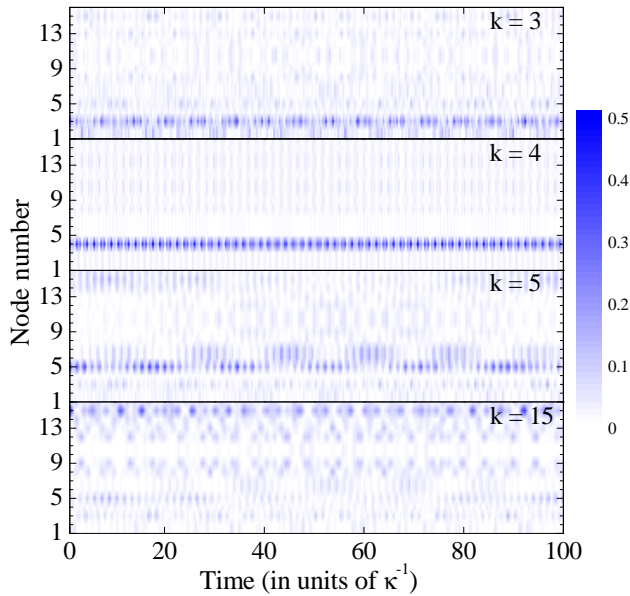}
\caption{\label{fig6} (Color online) Time evolution of the photonic excitation occupation probability, $\pi _{\ell k}^{ph}(t)$, 
through the $n=3$ AN (16 nodes) for the strong hopping regime. The vertical axis represents 
the occupation probability of a particular node (cavity).
The single cavity initial state
$\left| \psi_{k}(0) \right> = \left| k \right>\otimes (\left| g, 1 \right> 
+ \left| e, 0 \right>)/\sqrt{2}$
is prepared at nodes $k=3$, 4, 5, and 15. System parameters are $\kappa / \beta =10^{3}$, $\omega_{a}=0$, and $\omega_{f} = 0$. 
The node numbering is available in Fig. \ref{fig2}.}
\end{figure}

\begin{figure}[htb]
\includegraphics[width=0.47\textwidth]{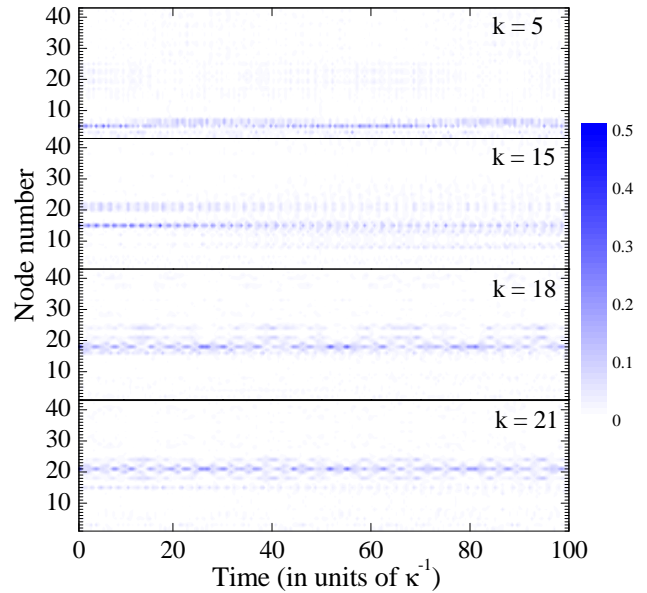}
\caption{\label{fig7} (Color online) Same as Fig. \ref{fig6}, but for $n=4$ (43 nodes) 
and $k=5$, 15, 18, and 21.}
\end{figure}

Firstly, observe that the AN has a $2\pi /3$ rotational symmetry, thus there are several 
subsets of equivalent nodes (see Fig. \ref{fig2}) like 
$\lbrace 1,2,3 \rbrace$, $\lbrace 5,6,7 \rbrace$, 
$\lbrace 8, 10, 11, 13, 14, 16 \rbrace$, and so forth, 
in which $\left| \psi _{k}(0) \right>$ would lead to the same dynamics. 
Likewise, for the initial state set at the hub node ($k=4$),
the occupation probabilities $\pi _{\ell 4}^{type}(t)$ are equal for each subset.
Figs. \ref{fig6} and \ref{fig7} 
show the time evolution of the photonic occupation probability 
for different initial single cavity
states on the 
$n=3$ and $n=4$ AN, respectively.
%
%
Although exhibiting a rather complex dynamics, we can identify some general
aspects of the propagation. The most remarkable feature is
periodic transition between localized and extended states. 
The return probability $\pi _{k k}^{ph}$ can reach the highest values so that  
the excitation is most likely to be found in the initial node.
This is expected due to the existence of many localized eigenstates, 
as discussed in Sec. \ref{sec3}. 
This localization is stronger for $k=4$,
the hub node. 
%
\begin{figure*}[t]
\includegraphics[width=0.95\textwidth]{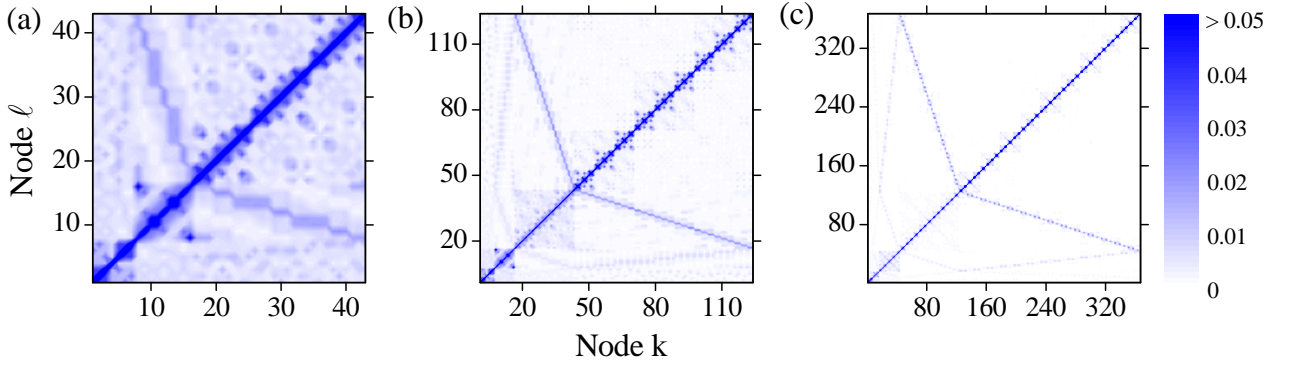}
\caption{\label{fig8}(Color online) Photonic long time average occupation probabilities $\chi_{\ell k}^{ph}$  
for (a) $n=4$ (43 nodes), (b) $n=5$ (124 nodes), and (c) $n=6$ (367 nodes). System parameters
are $\kappa / \beta = 10^{3}$, $\omega_{a} = 0$, and $\omega_{f}=0$. The plot shape is symmetrical since
$\chi_{\ell k}^{ph} = \chi_{k \ell}^{ph}$ 
for every $k$ and $\ell$, and the highest values are $\chi_{k k}^{ph}$,
i.e., the average return probability. Although we have not provided the node numbering for nodes from 
generations higher than $n=4$, the purpose here is to give an idea of how the subnetworks and nodes
from different generations are connected among themselves.
In each generation above, the darkest branch represents the average excitation transport probability
between nodes from the current generation and the previous one. The square boundaries
point out connections between nodes having the same $\gamma$.
}
\end{figure*} 
Now let us turn our attention to the probability of finding the excitation 
in nodes other than the initial one. In general, such probabilities are very small
compared to $\pi_{kk}^{ph}$, although these can reach relevant values in some specific nodes.
The probability distribution strongly depends on the given AN generation and the initial 
conditions.
In Fig. \ref{fig6}, by examining $\pi_{\ell k}^{ph}$ for $k=3$, we see that
a significant amount of probability flows to the other nodes belonging to the 
same degree group, that is, nodes 1 and 2. 
The same applies to $\pi_{65}^{ph}$ and $\pi_{75}^{ph}$ when $k=5$. For $k=15$, the nodes
10 and 11, even belonging to the same degree group, are barely populated. 
It suggests that sharing the same degree does not necessarily  
imply in privileged networking.
It turns clearer in Fig. \ref{fig7}, when $k=18$ and 21, where
the excitation mostly spreads to a few nodes
within the subnetwork 
\footnote{The AN has a self-similar structure where each subnetwork,
generated by 3 new corner nodes,
locally preserves the AN recurrence
but breaks the degree group symmetry of the original AN (see Fig. \ref{fig2}).} 
generated by the new corner nodes 1, 2, and 4. 
When $k=5$ for $n=4$, the excitation still roams between 
nodes 6 and 7, but now avoiding transport to higher generation nodes. For 
$k=15$, there is a relevant probability of finding 
the excitation at nodes 20, 21, and 22. These nodes belong
to the subnetwork formed by nodes 1, 2, and 5, where node 15 plays the role of hub node.
These observations indicate that these subnetworks have a fundamental part on the system
transport dynamics for higher AN generations. 

A better way to visualize how the excitation probability is distributed among the
nodes for given initial conditions is by evaluating the long time 
average of $\pi_{\ell k}^{ph}(t)$, defined as
\begin{eqnarray} \label{eqlongtime}
\chi_{\ell k}^{ph} &=& \lim_{T \rightarrow \infty} \dfrac{1}{T} \int_{0}^{T} \pi_{\ell k}^{ph}(t) dt \nonumber \\
&=& \sum_{i,j} \delta (E_{i}-E_{j}) \left< \ell, g \vert E_{i} \right>
\left< E_{i} \vert \psi _{k} (0) \right> \times \nonumber \\
 &\times & \left< \psi _{k} (0) \vert E_{j} \right> 
\left< E_{j} \vert \ell, g \right>,
\end{eqnarray}
where  $\delta (E_{i}-E_{j})=1$ for $E_{i}=E_{j}$ and $\delta (E_{i}-E_{j})=0$ else. 
Fig. \ref{fig8} shows the distribution of this average for different AN generations. 
First of all, notice that $\chi_{\ell k}^{ph} = \chi_{k \ell}^{ph}$ for every $k$ and $\ell$ (since
the evolution is unitary), hence this quantity provides us a clearer view on the way the nodes establish
communication channels with each other. As it must be, the higher averages are $\chi_{k k}^{ph}$, however,
by looking at the other connections we can identify 
several zones (square boundaries) and branches apart which, interestingly,
present self-similarity as $n$ increases.
The branches connect generations with each other while the zones connect
nodes from the same generation, i.e., with same degree $\gamma$. 
The strongest (darkest) branch
is always the one connecting nodes from the current generation to the previous one.
It mostly represents the links between the corresponding hub nodes from the $3^{n-2}$ 
subnetworks having 7 nodes, to its 
three nearest-neighbour nodes with $\gamma = 3$.    
As $n$ increases, the previous branches still remain but with lower intensities since
the aforementioned hub nodes now have a higher $\gamma$. 
Also, nodes from ``distant'' generations seem to be barely connected and  
links between nodes from the same generation are only relevant within subnetworks
composed by 16 nodes at most. For instance, at a given generation $n$, there are $3^{n-m}$ subnetworks
having $(3^{m}+5)/2$ nodes.
Further, these degree interconnections also become weaker as generation increases.  
    
\subsection{The JCH and strong atom-field coupling regimes}

In what follows we consider the situation where the hopping rate
and the atom-field coupling strength are of 
the same order: the JCH regime ($\kappa \approx \beta$). 
In this case the dynamics
is nontrivial since neither there is a JC-like interaction
between photonic and atomic components nor 
the photonic mode describes a quantum walk on the AN, as we have just seen for the strong hopping regime. Instead, somehow both degrees of
freedom start to interfere with each other. 

At the strong hopping regime it is possible to drive 
the atomic frequency (comprising a set of denegerate eigenvalues) along the JCH spectrum thus creating JC-like polaritons.
Otherwise we have well defined photonic- and atomic-like polaritons. 
Once the atom-field coupling parameter is raised, such effective resonance 
is no longer feasible. 
In other words,
at the JCH regime
it is not possible to generate a JC-like interaction
without disturb other polaritons associated with different
frequency levels on the JCH spectrum, within a range compatible 
with the $\kappa / \beta$ value. 
Thus, the excitation roams between photonic and atomic degrees 
of freedom while it propagates through the network in a rather complex way.
It turns out that each
cavity provides its contribution mostly to one 
of these two components, depending on the detuning and the
system initial conditions.
Hence, an initial state like 
$\left| \psi _{k}(0) \right> = \left| k \right>\otimes (\left| g, 1 \right> + 
\left| e, 0 \right>)/\sqrt{2}$ will not maintain the even superposition
between the whole photonic and atomic parts of the system,
as in the strong hopping case. 
Due to the existence of strongly localized polaritons,
we must expect that the detuning and the initial state
will be crucial in driving the whole system among both degrees of freedom.
%
\begin{figure*}[t!]
\includegraphics[width=0.8\textwidth]{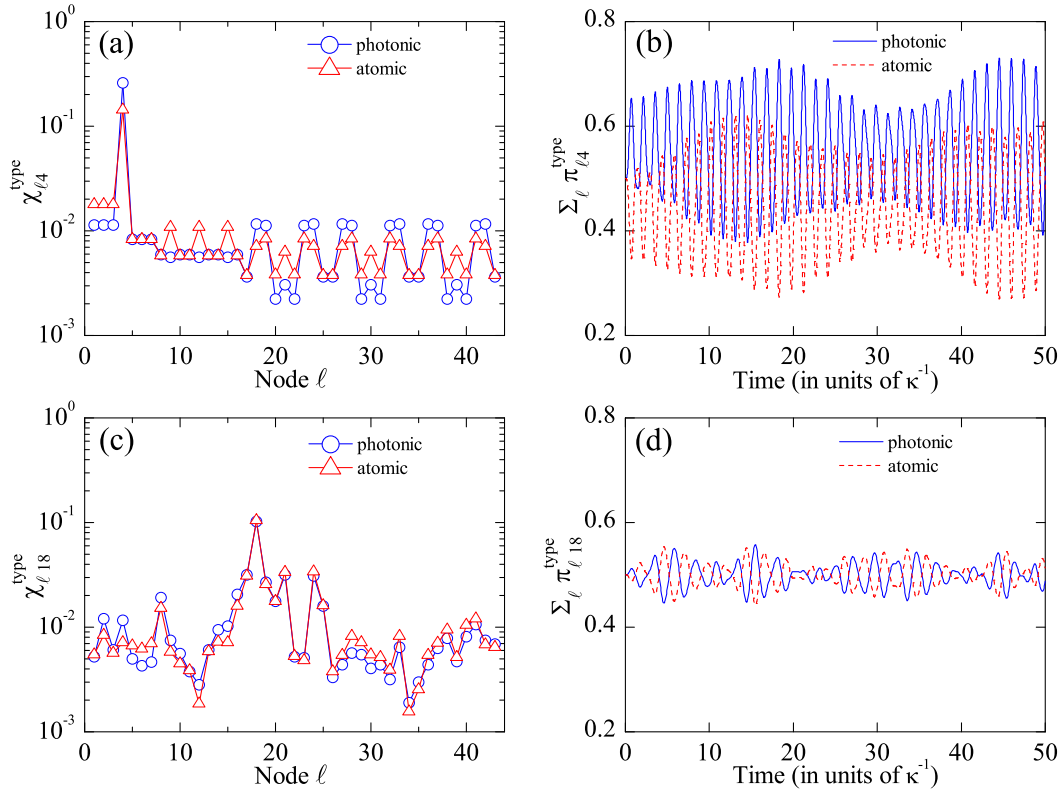}
\caption{\label{fig9}(Color online) Figures (a) and (c) show the 
photonic and atomic long time average 
occupation probability distribution for 
$\left| \psi _{k}(0) \right> = \left| k \right>\otimes (\left| g, 1 \right> + 
\left| e, 0 \right>)/\sqrt{2}$
prepared at
$k=4$ and $k=18$, respectively, and 
(b) and (d), the corresponding overall photonic and atomic overall probabilities. 
System parameters
are $n=4$, $\kappa / \beta = 1$, $\omega_{a} = 0$, and $\omega_{f} = 0$. 
The return average probabilities are
$\chi_{4 4}^{ph} \approx 0.26$, $\chi_{4 4}^{at} \approx 0.14$,
$\chi_{18 18}^{ph} \approx 0.103$, and $\chi_{18 18}^{at} \approx 0.105$.  
}
\end{figure*} 

Fig. \ref{fig9} describes how the initial state significantly
changes the way each cavity contributes to each one 
of the components, at the JCH regime. 
In Fig. \ref{fig9}(a) we show the 
long time average occupation probability distribution $\chi_{\ell k}^{type}$
when the single cavity state is set at the hub node, $k=4$. 
Notice that $\chi_{k \ell}^{at}$ can be evaluated by the same 
way as Eq. \ref{eqlongtime} by projecting the eigenstates 
on $\left| \ell, e \right>$, instead of $\left| \ell, g \right>$.
We can see that some groups of nodes have higher $\chi_{\ell 4}^{ph}$ values 
(including the hub node), others mostly contribute to $\chi_{\ell 4}^{at}$, 
and a few practically remains even for both components.
In Fig. \ref{fig9}(b) we
plot the 
overall photonic (atomic) occupation probability, 
$\sum_{\ell} \pi_{\ell k}^{ph}$ ($\sum_{\ell} \pi_{\ell k}^{at}$).
The whole system notoriously fluctuates between both components, mainly
due to large shift between $\chi_{4 4}^{ph}$ and $\chi_{4 4}^{at}$ 
plus the fact that the polaritons leading the dynamics are strongly
localized.  
Figs. \ref{fig9}(c) and \ref{fig9}(d) show the same quantities but now
for $k=18$ ($\gamma=3$). Recall that, as $\omega_{a}=0$, several
pairs of localized JC-like polaritons related to every node with $\gamma=3$ are
involved thus keeping $\chi_{18 18}^{ph}$ and $\chi_{18 18}^{at}$ practically
the same. However, even with a JC-like dynamics taking place, thus resulting
in a small variation of the overall occupation probability,
other
polaritons are involved, hence shifting the average for the rest of the cavities. 

Such nontrivial features arising from the JCH regime fade out if we keep
on raising the atom-field coupling strength until $\kappa \ll \beta$. 
In this regime, 
the dynamics for both atomic and photonic modes 
behaves exactly like at the large hopping regime but now creating new single cavity eigenstates
(if $\left| \psi _{k}(0) \right> = \left| k \right>\otimes (\left| g, 1 \right> + 
\left| e, 0 \right>)/\sqrt{2}$), 
as passing by other cavities, and propagating at half the speed
of the photonic excitation at the $\kappa \gg \beta$ regime \cite{makin09}.
It turns out to be expected, from the fact that, as $\kappa / \beta$ decreases,
more JC-like pairs of polaritons are created until the entire JCH spectrum is affected.

\subsection{Results overview}

%
\begin{figure*}[t!]
\includegraphics[width=0.75\textwidth]{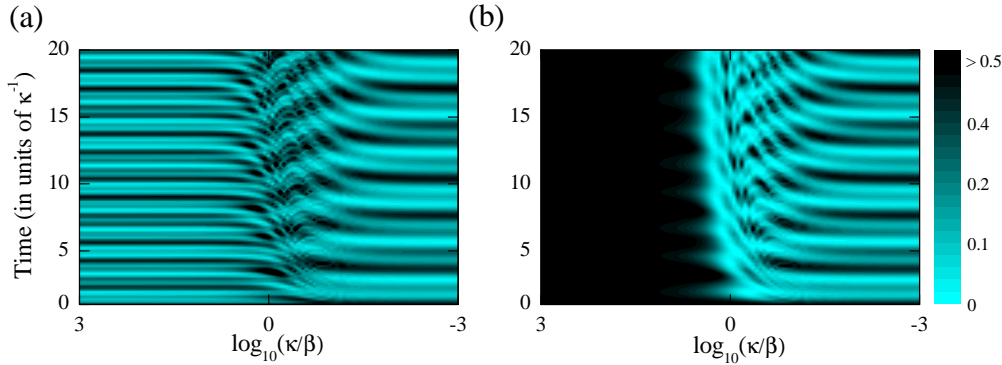}
\caption{\label{fig10} (Color online) Comparison between atomic and photonic propagation
dynamics from the large hopping regime, $\kappa / \beta = 10^{3}$, to the strong
atom-field coupling regime, $\kappa / \beta = 10^{-3}$, for an initial
single cavity state 
$\left| \psi _{k}(0) \right> = \left| k \right>\otimes (\left| g, 1 \right> + 
\left| e, 0 \right>)/\sqrt{2}$ prepared at $k=4$. 
In (a), the photonic return 
probability, $\pi_{44}^{ph}$, and in (b), the atomic counterpart. System  
parameters are $n=4$, $\omega_{a} = 0$, and $\omega_{f} = 0$. The middle range of $\kappa / \beta$ values is where the JCH
regime overcomes. 
}
\end{figure*} 

We now provide a review of the main features provided by the different
energy regimes, as previously discussed. Although we have dealt with the JCH
dynamics on the AN, these properties apply for any structure. The main
difference lies on the existence of a gapped spectrum and localized eigenstates, and
also whether or not it holds for a large-size system.

In Fig. \ref{fig10} we
outline the energy regime transition from $\kappa / \beta = 10^{3}$ to
$10^{-3}$ by evaluating the  
time evolution of the photonic [Fig. \ref{fig10}(a)] 
and atomic [Fig. \ref{fig10}(b)] return probabilities $\pi_{44}^{type}$ for
a single cavity initial state, Eq. (\ref{scstate}).
At the strong hopping regime, $\kappa \gg \beta$, the interaction 
between photonic and atomic components
occurs in a time scale much longer than the photon hopping. 
It means that, in a single cavity initial state,
the whole system remains in an even superposition between both components.
As the single photon describes a quantum walk through the network,
the atom-field coupling does not interfere on it, practically. Moreover, the atomic dynamics
will be restricted to a single or a set of JC-like pair of polaritons (if the corresponding energy level is degenerate), 
or none, if the atomic frequency $\omega_{a}$ does not match with any field normal mode frequency. The latter case
obviously implies in the atomic freezing. On the contrary, a JC-like pair (or pairs) of polaritons might 
be created and, if it happens that it is strongly localized at some node, by setting the appropriate
initial state, a significant atom-field energy exchange takes place 
(in a time scale of the order of $\beta$), as described in
Figs. \ref{fig4} and \ref{fig5}, where we have considered a fully atomic initial state in order to highlight 
the phenomena. 
We have also analysed the quantum transport properties of the AN.
The excitation is most likely to be found at the 
initial node, as expected.
By evaluating the long time average photonic occupation probability
for different AN generations,
we have discussed the interplay between the nodes
and subnetworks thus showing their role on the system dynamics.
At a given generation $n$, the nodes with $\gamma=3$ are mostly connected with its 
correspondent hub node from the previous generation and with some other nodes
having the same degree within local subnetworks. Furthermore, these properties hold
recurrently as we increase the network size (see Fig. \ref{fig8}).

The JCH regime, $\kappa \approx \beta$, is characterized by a polariton interference which
leads the system to roam between photonic and atomic degrees of freedom 
in a nontrivial way. The dynamics is neither strictly ruled by JC-like pairs of polaritons nor the photonic mode propagates freely. Instead, for a given
detuning, even if JC-like pairs of polaritons are set, other polaritons
associated with different frequency levels on the JCH spectrum now
contribute to the dynamics.
It generates a splitting between photonic and atomic average occupation
(Fig. \ref{fig9}) within each cavity.
Such deviation can be avoided (at least in a particular cavity) 
by setting an appropriate atomic frequency in order to create
a pair of JC-like polaritons strongly localized at the initial node.

The strong atom-field coupling regime, $\kappa \ll \beta $, is reached by
the time such polariton interference is so robust that every field normal
mode becomes coupled to its atomic analog and hence the photonic and
atomic excitation propagate coherently through the network.

\section{\label{sec5}Conclusions}

Even at the single excitation subspace, the JCH model provides 
a rich dynamics. Moreover,
the AN induces a peculiar spectrum
and set of strongly localized eigenstates
due solely to its topology, as we have considered uniform 
coupling parameters over the network. The structure self-similarity
naturally reflects on the JCH model dynamics. As a result,
the atomic excitation might not be stationary at the strong hopping
regime even for large-size networks. Another appealing feature
is the possibility of driving a specific set of polaritons
to rule the dynamics,
by a judicious tuning of the atomic transition frequency.

We have also discussed
the continuous time quantum walk on the AN for different system sizes. 
For any initial state (fully localized at a specific node), the
system will always carry a large amount of information about its initial conditions
and will significantly propagate to a small set of nodes only.
Further work along this direction should also consider other
kinds of initial states, other than localized, where a much
more complex dynamics arise.

Even though high-fidelity excitation transport protocols 
in an one-dimensional array of 
coupled cavities have been proposed \cite{makin09, dong12}, such regular structures 
induce mostly extended polaritons,
which do not highlight all the features
the single excitation JCH model can provide, in contrast 
to non-conventional complex structures, where, aside from the JCH model,
might also shed light in the behavior of other many-body systems.
 
\section*{Acknowledgements}

Fruitful discussions with F. Ciccarello are gratefully acknowledged. 
A.M.C.S. wishes to thank the National Institute of
Science and Technology for Complex Systems.
This work was supported by the CNPq (Brazilian agency).


\begin{thebibliography}{42}%
\makeatletter
\providecommand \@ifxundefined [1]{%
 \@ifx{#1\undefined}
}%
\providecommand \@ifnum [1]{%
 \ifnum #1\expandafter \@firstoftwo
 \else \expandafter \@secondoftwo
 \fi
}%
\providecommand \@ifx [1]{%
 \ifx #1\expandafter \@firstoftwo
 \else \expandafter \@secondoftwo
 \fi
}%
\providecommand \natexlab [1]{#1}%
\providecommand \enquote  [1]{``#1''}%
\providecommand \bibnamefont  [1]{#1}%
\providecommand \bibfnamefont [1]{#1}%
\providecommand \citenamefont [1]{#1}%
\providecommand \href@noop [0]{\@secondoftwo}%
\providecommand \href [0]{\begingroup \@sanitize@url \@href}%
\providecommand \@href[1]{\@@startlink{#1}\@@href}%
\providecommand \@@href[1]{\endgroup#1\@@endlink}%
\providecommand \@sanitize@url [0]{\catcode `\\12\catcode `\$12\catcode
  `\&12\catcode `\#12\catcode `\^12\catcode `\_12\catcode `\%12\relax}%
\providecommand \@@startlink[1]{}%
\providecommand \@@endlink[0]{}%
\providecommand \url  [0]{\begingroup\@sanitize@url \@url }%
\providecommand \@url [1]{\endgroup\@href {#1}{\urlprefix }}%
\providecommand \urlprefix  [0]{URL }%
\providecommand \Eprint [0]{\href }%
\providecommand \doibase [0]{http://dx.doi.org/}%
\providecommand \selectlanguage [0]{\@gobble}%
\providecommand \bibinfo  [0]{\@secondoftwo}%
\providecommand \bibfield  [0]{\@secondoftwo}%
\providecommand \translation [1]{[#1]}%
\providecommand \BibitemOpen [0]{}%
\providecommand \bibitemStop [0]{}%
\providecommand \bibitemNoStop [0]{.\EOS\space}%
\providecommand \EOS [0]{\spacefactor3000\relax}%
\providecommand \BibitemShut  [1]{\csname bibitem#1\endcsname}%
\let\auto@bib@innerbib\@empty
\bibitem [{\citenamefont {Nielsen}\ and\ \citenamefont
  {Chuang}(2000)}]{nielsen}%
  \BibitemOpen
  \bibfield  {author} {\bibinfo {author} {\bibfnamefont {M.~A.}\ \bibnamefont
  {Nielsen}}\ and\ \bibinfo {author} {\bibfnamefont {I.~L.}\ \bibnamefont
  {Chuang}},\ }\href@noop {} {\emph {\bibinfo {title} {Quantum Computation and
  Quantum Information}}}\ (\bibinfo  {publisher} {Cambridge University Press},\
  \bibinfo {year} {2000})\BibitemShut {NoStop}%
\bibitem [{\citenamefont {Pachos}\ and\ \citenamefont
  {Walther}(2002)}]{pachos02}%
  \BibitemOpen
  \bibfield  {author} {\bibinfo {author} {\bibfnamefont {J.}~\bibnamefont
  {Pachos}}\ and\ \bibinfo {author} {\bibfnamefont {H.}~\bibnamefont
  {Walther}},\ }\href {\doibase 10.1103/PhysRevLett.89.187903} {\bibfield
  {journal} {\bibinfo  {journal} {Phys. Rev. Lett.}\ }\textbf {\bibinfo
  {volume} {89}},\ \bibinfo {pages} {187903} (\bibinfo {year}
  {2002})}\BibitemShut {NoStop}%
\bibitem [{\citenamefont {Blais}\ \emph {et~al.}(2004)\citenamefont {Blais},
  \citenamefont {Huang}, \citenamefont {Wallraff}, \citenamefont {Girvin},\
  and\ \citenamefont {Schoelkopf}}]{blais04}%
  \BibitemOpen
  \bibfield  {author} {\bibinfo {author} {\bibfnamefont {A.}~\bibnamefont
  {Blais}}, \bibinfo {author} {\bibfnamefont {R.-S.}\ \bibnamefont {Huang}},
  \bibinfo {author} {\bibfnamefont {A.}~\bibnamefont {Wallraff}}, \bibinfo
  {author} {\bibfnamefont {S.~M.}\ \bibnamefont {Girvin}}, \ and\ \bibinfo
  {author} {\bibfnamefont {R.~J.}\ \bibnamefont {Schoelkopf}},\ }\href
  {\doibase 10.1103/PhysRevA.69.062320} {\bibfield  {journal} {\bibinfo
  {journal} {Phys. Rev. A}\ }\textbf {\bibinfo {volume} {69}},\ \bibinfo
  {pages} {062320} (\bibinfo {year} {2004})}\BibitemShut {NoStop}%
\bibitem [{\citenamefont {Su}\ \emph {et~al.}(2008)\citenamefont {Su},
  \citenamefont {Greentree}, \citenamefont {Munro}, \citenamefont {Nemoto},\
  and\ \citenamefont {Hollenberg}}]{su08}%
  \BibitemOpen
  \bibfield  {author} {\bibinfo {author} {\bibfnamefont {C.-H.}\ \bibnamefont
  {Su}}, \bibinfo {author} {\bibfnamefont {A.~D.}\ \bibnamefont {Greentree}},
  \bibinfo {author} {\bibfnamefont {W.~J.}\ \bibnamefont {Munro}}, \bibinfo
  {author} {\bibfnamefont {K.}~\bibnamefont {Nemoto}}, \ and\ \bibinfo {author}
  {\bibfnamefont {L.~C.~L.}\ \bibnamefont {Hollenberg}},\ }\href {\doibase
  10.1103/PhysRevA.78.062336} {\bibfield  {journal} {\bibinfo  {journal} {Phys.
  Rev. A}\ }\textbf {\bibinfo {volume} {78}},\ \bibinfo {pages} {062336}
  (\bibinfo {year} {2008})}\BibitemShut {NoStop}%
\bibitem [{\citenamefont {Jaynes}\ and\ \citenamefont
  {Cummings}(1963)}]{jaynes63}%
  \BibitemOpen
  \bibfield  {author} {\bibinfo {author} {\bibfnamefont {E.~T.}\ \bibnamefont
  {Jaynes}}\ and\ \bibinfo {author} {\bibfnamefont {F.~W.}\ \bibnamefont
  {Cummings}},\ }in\ \href@noop {} {\emph {\bibinfo {booktitle} {Proc. of the
  IEEE}}},\ Vol.~\bibinfo {volume} {51}\ (\bibinfo {year} {1963})\ p.~\bibinfo
  {pages} {89}\BibitemShut {NoStop}%
\bibitem [{\citenamefont {Shore}\ and\ \citenamefont {Knight}(1993)}]{shore93}%
  \BibitemOpen
  \bibfield  {author} {\bibinfo {author} {\bibfnamefont {B.~W.}\ \bibnamefont
  {Shore}}\ and\ \bibinfo {author} {\bibfnamefont {P.~L.}\ \bibnamefont
  {Knight}},\ }\href@noop {} {\bibfield  {journal} {\bibinfo  {journal} {J.
  Mod. Optics}\ }\textbf {\bibinfo {volume} {40}},\ \bibinfo {pages} {1195}
  (\bibinfo {year} {1993})}\BibitemShut {NoStop}%
\bibitem [{\citenamefont {Armani}\ \emph {et~al.}(2003)\citenamefont {Armani},
  \citenamefont {Kippenberg}, \citenamefont {Spillane},\ and\ \citenamefont
  {Vahala}}]{armani03}%
  \BibitemOpen
  \bibfield  {author} {\bibinfo {author} {\bibfnamefont {D.~K.}\ \bibnamefont
  {Armani}}, \bibinfo {author} {\bibfnamefont {T.~J.}\ \bibnamefont
  {Kippenberg}}, \bibinfo {author} {\bibfnamefont {S.~M.}\ \bibnamefont
  {Spillane}}, \ and\ \bibinfo {author} {\bibfnamefont {K.~J.}\ \bibnamefont
  {Vahala}},\ }\href@noop {} {\bibfield  {journal} {\bibinfo  {journal} {Nature
  (London)}\ }\textbf {\bibinfo {volume} {421}},\ \bibinfo {pages} {925}
  (\bibinfo {year} {2003})}\BibitemShut {NoStop}%
\bibitem [{\citenamefont {Hennessy}\ \emph {et~al.}(2007)\citenamefont
  {Hennessy}, \citenamefont {Badolato}, \citenamefont {Winger}, \citenamefont
  {Gerace}, \citenamefont {Atatue}, \citenamefont {Gulde}, \citenamefont
  {Falt}, \citenamefont {Hu},\ and\ \citenamefont {Imamoglu}}]{hennessy07}%
  \BibitemOpen
  \bibfield  {author} {\bibinfo {author} {\bibfnamefont {K.}~\bibnamefont
  {Hennessy}}, \bibinfo {author} {\bibfnamefont {A.}~\bibnamefont {Badolato}},
  \bibinfo {author} {\bibfnamefont {M.}~\bibnamefont {Winger}}, \bibinfo
  {author} {\bibfnamefont {D.}~\bibnamefont {Gerace}}, \bibinfo {author}
  {\bibfnamefont {M.}~\bibnamefont {Atatue}}, \bibinfo {author} {\bibfnamefont
  {S.}~\bibnamefont {Gulde}}, \bibinfo {author} {\bibfnamefont
  {S.}~\bibnamefont {Falt}}, \bibinfo {author} {\bibfnamefont {E.~L.}\
  \bibnamefont {Hu}}, \ and\ \bibinfo {author} {\bibfnamefont {A.}~\bibnamefont
  {Imamoglu}},\ }\href@noop {} {\bibfield  {journal} {\bibinfo  {journal}
  {Nature (London)}\ }\textbf {\bibinfo {volume} {445}},\ \bibinfo {pages}
  {896} (\bibinfo {year} {2007})}\BibitemShut {NoStop}%
\bibitem [{\citenamefont {Wallraff}\ \emph {et~al.}(2004)\citenamefont
  {Wallraff}, \citenamefont {Schuster}, \citenamefont {Blais}, \citenamefont
  {Frunzio}, \citenamefont {Huang}, \citenamefont {Majer}, \citenamefont
  {Kumar}, \citenamefont {Girvin},\ and\ \citenamefont
  {Schoelkopf}}]{wallraff04}%
  \BibitemOpen
  \bibfield  {author} {\bibinfo {author} {\bibfnamefont {A.}~\bibnamefont
  {Wallraff}}, \bibinfo {author} {\bibfnamefont {D.~I.}\ \bibnamefont
  {Schuster}}, \bibinfo {author} {\bibfnamefont {A.}~\bibnamefont {Blais}},
  \bibinfo {author} {\bibfnamefont {L.}~\bibnamefont {Frunzio}}, \bibinfo
  {author} {\bibfnamefont {R.~S.}\ \bibnamefont {Huang}}, \bibinfo {author}
  {\bibfnamefont {J.}~\bibnamefont {Majer}}, \bibinfo {author} {\bibfnamefont
  {S.}~\bibnamefont {Kumar}}, \bibinfo {author} {\bibfnamefont {S.~M.}\
  \bibnamefont {Girvin}}, \ and\ \bibinfo {author} {\bibfnamefont {R.~J.}\
  \bibnamefont {Schoelkopf}},\ }\href@noop {} {\bibfield  {journal} {\bibinfo
  {journal} {Nature (London)}\ }\textbf {\bibinfo {volume} {431}},\ \bibinfo
  {pages} {162} (\bibinfo {year} {2004})}\BibitemShut {NoStop}%
\bibitem [{\citenamefont {Hartmann}\ \emph {et~al.}(2008)\citenamefont
  {Hartmann}, \citenamefont {Brand\~{a}o},\ and\ \citenamefont
  {Plenio}}]{hartmann08rev}%
  \BibitemOpen
  \bibfield  {author} {\bibinfo {author} {\bibfnamefont {M.~J.}\ \bibnamefont
  {Hartmann}}, \bibinfo {author} {\bibfnamefont {F.~G. S.~L.}\ \bibnamefont
  {Brand\~{a}o}}, \ and\ \bibinfo {author} {\bibfnamefont {M.~B.}\ \bibnamefont
  {Plenio}},\ }\href@noop {} {\bibfield  {journal} {\bibinfo  {journal} {Laser
  Photon. Rev.}\ }\textbf {\bibinfo {volume} {2}},\ \bibinfo {pages} {527}
  (\bibinfo {year} {2008})}\BibitemShut {NoStop}%
\bibitem [{\citenamefont {Tomadin}\ and\ \citenamefont
  {Fazio}(2010)}]{tomadin10}%
  \BibitemOpen
  \bibfield  {author} {\bibinfo {author} {\bibfnamefont {A.}~\bibnamefont
  {Tomadin}}\ and\ \bibinfo {author} {\bibfnamefont {R.}~\bibnamefont
  {Fazio}},\ }\href@noop {} {\bibfield  {journal} {\bibinfo  {journal} {J. Opt.
  Soc. Am. B}\ }\textbf {\bibinfo {volume} {27}},\ \bibinfo {pages} {A130}
  (\bibinfo {year} {2010})}\BibitemShut {NoStop}%
\bibitem [{\citenamefont {Greentree}\ \emph {et~al.}(2006)\citenamefont
  {Greentree}, \citenamefont {Tahan}, \citenamefont {Cole},\ and\ \citenamefont
  {Hollenberg}}]{greentree06}%
  \BibitemOpen
  \bibfield  {author} {\bibinfo {author} {\bibfnamefont {A.~D.}\ \bibnamefont
  {Greentree}}, \bibinfo {author} {\bibfnamefont {C.}~\bibnamefont {Tahan}},
  \bibinfo {author} {\bibfnamefont {J.~H.}\ \bibnamefont {Cole}}, \ and\
  \bibinfo {author} {\bibfnamefont {L.~C.~L.}\ \bibnamefont {Hollenberg}},\
  }\href@noop {} {\bibfield  {journal} {\bibinfo  {journal} {Nat. Phys.}\
  }\textbf {\bibinfo {volume} {2}},\ \bibinfo {pages} {856} (\bibinfo {year}
  {2006})}\BibitemShut {NoStop}%
\bibitem [{\citenamefont {Hartmann}\ \emph {et~al.}(2006)\citenamefont
  {Hartmann}, \citenamefont {Brand\~{a}o},\ and\ \citenamefont
  {Plenio}}]{hartmann06}%
  \BibitemOpen
  \bibfield  {author} {\bibinfo {author} {\bibfnamefont {M.~J.}\ \bibnamefont
  {Hartmann}}, \bibinfo {author} {\bibfnamefont {F.~G. S.~L.}\ \bibnamefont
  {Brand\~{a}o}}, \ and\ \bibinfo {author} {\bibfnamefont {M.~B.}\ \bibnamefont
  {Plenio}},\ }\href@noop {} {\bibfield  {journal} {\bibinfo  {journal} {Nat.
  Phys.}\ }\textbf {\bibinfo {volume} {2}},\ \bibinfo {pages} {849} (\bibinfo
  {year} {2006})}\BibitemShut {NoStop}%
\bibitem [{\citenamefont {Angelakis}\ \emph {et~al.}(2007)\citenamefont
  {Angelakis}, \citenamefont {Santos},\ and\ \citenamefont
  {Bose}}]{angelakis07}%
  \BibitemOpen
  \bibfield  {author} {\bibinfo {author} {\bibfnamefont {D.~G.}\ \bibnamefont
  {Angelakis}}, \bibinfo {author} {\bibfnamefont {M.~F.}\ \bibnamefont
  {Santos}}, \ and\ \bibinfo {author} {\bibfnamefont {S.}~\bibnamefont
  {Bose}},\ }\href {\doibase 10.1103/PhysRevA.76.031805} {\bibfield  {journal}
  {\bibinfo  {journal} {Phys. Rev. A}\ }\textbf {\bibinfo {volume} {76}},\
  \bibinfo {pages} {031805(R)} (\bibinfo {year} {2007})}\BibitemShut {NoStop}%
\bibitem [{\citenamefont {Rossini}\ and\ \citenamefont
  {Fazio}(2007)}]{rossini07}%
  \BibitemOpen
  \bibfield  {author} {\bibinfo {author} {\bibfnamefont {D.}~\bibnamefont
  {Rossini}}\ and\ \bibinfo {author} {\bibfnamefont {R.}~\bibnamefont
  {Fazio}},\ }\href {\doibase 10.1103/PhysRevLett.99.186401} {\bibfield
  {journal} {\bibinfo  {journal} {Phys. Rev. Lett.}\ }\textbf {\bibinfo
  {volume} {99}},\ \bibinfo {pages} {186401} (\bibinfo {year}
  {2007})}\BibitemShut {NoStop}%
\bibitem [{\citenamefont {Aichhorn}\ \emph {et~al.}(2008)\citenamefont
  {Aichhorn}, \citenamefont {Hohenadler}, \citenamefont {Tahan},\ and\
  \citenamefont {Littlewood}}]{aichhorn08}%
  \BibitemOpen
  \bibfield  {author} {\bibinfo {author} {\bibfnamefont {M.}~\bibnamefont
  {Aichhorn}}, \bibinfo {author} {\bibfnamefont {M.}~\bibnamefont
  {Hohenadler}}, \bibinfo {author} {\bibfnamefont {C.}~\bibnamefont {Tahan}}, \
  and\ \bibinfo {author} {\bibfnamefont {P.~B.}\ \bibnamefont {Littlewood}},\
  }\href {\doibase 10.1103/PhysRevLett.100.216401} {\bibfield  {journal}
  {\bibinfo  {journal} {Phys. Rev. Lett.}\ }\textbf {\bibinfo {volume} {100}},\
  \bibinfo {pages} {216401} (\bibinfo {year} {2008})}\BibitemShut {NoStop}%
\bibitem [{\citenamefont {Koch}\ and\ \citenamefont {Le~Hur}(2009)}]{lehur09}%
  \BibitemOpen
  \bibfield  {author} {\bibinfo {author} {\bibfnamefont {J.}~\bibnamefont
  {Koch}}\ and\ \bibinfo {author} {\bibfnamefont {K.}~\bibnamefont {Le~Hur}},\
  }\href {\doibase 10.1103/PhysRevA.80.023811} {\bibfield  {journal} {\bibinfo
  {journal} {Phys. Rev. A}\ }\textbf {\bibinfo {volume} {80}},\ \bibinfo
  {pages} {023811} (\bibinfo {year} {2009})}\BibitemShut {NoStop}%
\bibitem [{\citenamefont {Pippan}\ \emph {et~al.}(2009)\citenamefont {Pippan},
  \citenamefont {Evertz},\ and\ \citenamefont {Hohenadler}}]{pippan09}%
  \BibitemOpen
  \bibfield  {author} {\bibinfo {author} {\bibfnamefont {P.}~\bibnamefont
  {Pippan}}, \bibinfo {author} {\bibfnamefont {H.~G.}\ \bibnamefont {Evertz}},
  \ and\ \bibinfo {author} {\bibfnamefont {M.}~\bibnamefont {Hohenadler}},\
  }\href {\doibase 10.1103/PhysRevA.80.033612} {\bibfield  {journal} {\bibinfo
  {journal} {Phys. Rev. A}\ }\textbf {\bibinfo {volume} {80}},\ \bibinfo
  {pages} {033612} (\bibinfo {year} {2009})}\BibitemShut {NoStop}%
\bibitem [{\citenamefont {Schmidt}\ and\ \citenamefont
  {Blatter}(2009)}]{schmidt09}%
  \BibitemOpen
  \bibfield  {author} {\bibinfo {author} {\bibfnamefont {S.}~\bibnamefont
  {Schmidt}}\ and\ \bibinfo {author} {\bibfnamefont {G.}~\bibnamefont
  {Blatter}},\ }\href {\doibase 10.1103/PhysRevLett.103.086403} {\bibfield
  {journal} {\bibinfo  {journal} {Phys. Rev. Lett.}\ }\textbf {\bibinfo
  {volume} {103}},\ \bibinfo {pages} {086403} (\bibinfo {year}
  {2009})}\BibitemShut {NoStop}%
\bibitem [{\citenamefont {Knap}\ \emph {et~al.}(2010)\citenamefont {Knap},
  \citenamefont {Arrigoni},\ and\ \citenamefont {von~der Linden}}]{knap10}%
  \BibitemOpen
  \bibfield  {author} {\bibinfo {author} {\bibfnamefont {M.}~\bibnamefont
  {Knap}}, \bibinfo {author} {\bibfnamefont {E.}~\bibnamefont {Arrigoni}}, \
  and\ \bibinfo {author} {\bibfnamefont {W.}~\bibnamefont {von~der Linden}},\
  }\href {\doibase 10.1103/PhysRevB.81.104303} {\bibfield  {journal} {\bibinfo
  {journal} {Phys. Rev. B}\ }\textbf {\bibinfo {volume} {81}},\ \bibinfo
  {pages} {104303} (\bibinfo {year} {2010})}\BibitemShut {NoStop}%
\bibitem [{\citenamefont {Hohenadler}\ \emph {et~al.}(2011)\citenamefont
  {Hohenadler}, \citenamefont {Aichhorn}, \citenamefont {Schmidt},\ and\
  \citenamefont {Pollet}}]{hohenadler11}%
  \BibitemOpen
  \bibfield  {author} {\bibinfo {author} {\bibfnamefont {M.}~\bibnamefont
  {Hohenadler}}, \bibinfo {author} {\bibfnamefont {M.}~\bibnamefont
  {Aichhorn}}, \bibinfo {author} {\bibfnamefont {S.}~\bibnamefont {Schmidt}}, \
  and\ \bibinfo {author} {\bibfnamefont {L.}~\bibnamefont {Pollet}},\ }\href
  {\doibase 10.1103/PhysRevA.84.041608} {\bibfield  {journal} {\bibinfo
  {journal} {Phys. Rev. A}\ }\textbf {\bibinfo {volume} {84}},\ \bibinfo
  {pages} {041608} (\bibinfo {year} {2011})}\BibitemShut {NoStop}%
\bibitem [{\citenamefont {Cirac}\ \emph {et~al.}(1999)\citenamefont {Cirac},
  \citenamefont {Ekert}, \citenamefont {Huelga},\ and\ \citenamefont
  {Macchiavello}}]{cirac99}%
  \BibitemOpen
  \bibfield  {author} {\bibinfo {author} {\bibfnamefont {J.~I.}\ \bibnamefont
  {Cirac}}, \bibinfo {author} {\bibfnamefont {A.~K.}\ \bibnamefont {Ekert}},
  \bibinfo {author} {\bibfnamefont {S.~F.}\ \bibnamefont {Huelga}}, \ and\
  \bibinfo {author} {\bibfnamefont {C.}~\bibnamefont {Macchiavello}},\
  }\href@noop {} {\bibfield  {journal} {\bibinfo  {journal} {Phys. Rev. A}\
  }\textbf {\bibinfo {volume} {59}},\ \bibinfo {pages} {4249} (\bibinfo {year}
  {1999})}\BibitemShut {NoStop}%
\bibitem [{\citenamefont {Angelakis}\ and\ \citenamefont
  {Bose}(2007)}]{angelakis07-ent}%
  \BibitemOpen
  \bibfield  {author} {\bibinfo {author} {\bibfnamefont {D.~G.}\ \bibnamefont
  {Angelakis}}\ and\ \bibinfo {author} {\bibfnamefont {S.}~\bibnamefont
  {Bose}},\ }\href@noop {} {\bibfield  {journal} {\bibinfo  {journal} {J. Opt.
  Soc. Am. B}\ }\textbf {\bibinfo {volume} {24}},\ \bibinfo {pages} {266}
  (\bibinfo {year} {2007})}\BibitemShut {NoStop}%
\bibitem [{\citenamefont {Cirac}\ \emph {et~al.}(1997)\citenamefont {Cirac},
  \citenamefont {Zoller}, \citenamefont {Kimble},\ and\ \citenamefont
  {Mabuchi}}]{cirac97}%
  \BibitemOpen
  \bibfield  {author} {\bibinfo {author} {\bibfnamefont {J.~I.}\ \bibnamefont
  {Cirac}}, \bibinfo {author} {\bibfnamefont {P.}~\bibnamefont {Zoller}},
  \bibinfo {author} {\bibfnamefont {H.~J.}\ \bibnamefont {Kimble}}, \ and\
  \bibinfo {author} {\bibfnamefont {H.}~\bibnamefont {Mabuchi}},\ }\href@noop
  {} {\bibfield  {journal} {\bibinfo  {journal} {Phys. Rev. Lett.}\ }\textbf
  {\bibinfo {volume} {78}},\ \bibinfo {pages} {3221} (\bibinfo {year}
  {1997})}\BibitemShut {NoStop}%
\bibitem [{\citenamefont {Bose}\ \emph {et~al.}(2007)\citenamefont {Bose},
  \citenamefont {Angelakis},\ and\ \citenamefont {Burgarth}}]{bose07-qst}%
  \BibitemOpen
  \bibfield  {author} {\bibinfo {author} {\bibfnamefont {S.}~\bibnamefont
  {Bose}}, \bibinfo {author} {\bibfnamefont {D.~G.}\ \bibnamefont {Angelakis}},
  \ and\ \bibinfo {author} {\bibfnamefont {D.}~\bibnamefont {Burgarth}},\
  }\href@noop {} {\bibfield  {journal} {\bibinfo  {journal} {J. Mod. Opt.}\
  }\textbf {\bibinfo {volume} {54}},\ \bibinfo {pages} {2307} (\bibinfo {year}
  {2007})}\BibitemShut {NoStop}%
\bibitem [{\citenamefont {Ogden}\ \emph {et~al.}(2008)\citenamefont {Ogden},
  \citenamefont {Irish},\ and\ \citenamefont {Kim}}]{ogden08}%
  \BibitemOpen
  \bibfield  {author} {\bibinfo {author} {\bibfnamefont {C.~D.}\ \bibnamefont
  {Ogden}}, \bibinfo {author} {\bibfnamefont {E.~K.}\ \bibnamefont {Irish}}, \
  and\ \bibinfo {author} {\bibfnamefont {M.~S.}\ \bibnamefont {Kim}},\ }\href
  {\doibase 10.1103/PhysRevA.78.063805} {\bibfield  {journal} {\bibinfo
  {journal} {Phys. Rev. A}\ }\textbf {\bibinfo {volume} {78}},\ \bibinfo
  {pages} {063805} (\bibinfo {year} {2008})}\BibitemShut {NoStop}%
\bibitem [{\citenamefont {Makin}\ \emph {et~al.}(2009)\citenamefont {Makin},
  \citenamefont {Cole}, \citenamefont {Hill}, \citenamefont {Greentree},\ and\
  \citenamefont {Hollenberg}}]{makin09}%
  \BibitemOpen
  \bibfield  {author} {\bibinfo {author} {\bibfnamefont {M.~I.}\ \bibnamefont
  {Makin}}, \bibinfo {author} {\bibfnamefont {J.~H.}\ \bibnamefont {Cole}},
  \bibinfo {author} {\bibfnamefont {C.~D.}\ \bibnamefont {Hill}}, \bibinfo
  {author} {\bibfnamefont {A.~D.}\ \bibnamefont {Greentree}}, \ and\ \bibinfo
  {author} {\bibfnamefont {L.~C.~L.}\ \bibnamefont {Hollenberg}},\ }\href
  {\doibase 10.1103/PhysRevA.80.043842} {\bibfield  {journal} {\bibinfo
  {journal} {Phys. Rev. A}\ }\textbf {\bibinfo {volume} {80}},\ \bibinfo
  {pages} {043842} (\bibinfo {year} {2009})}\BibitemShut {NoStop}%
\bibitem [{\citenamefont {Ciccarello}(2011)}]{ciccarello11}%
  \BibitemOpen
  \bibfield  {author} {\bibinfo {author} {\bibfnamefont {F.}~\bibnamefont
  {Ciccarello}},\ }\href {\doibase 10.1103/PhysRevA.83.043802} {\bibfield
  {journal} {\bibinfo  {journal} {Phys. Rev. A}\ }\textbf {\bibinfo {volume}
  {83}},\ \bibinfo {pages} {043802} (\bibinfo {year} {2011})}\BibitemShut
  {NoStop}%
\bibitem [{\citenamefont {Dong}\ \emph {et~al.}(2012)\citenamefont {Dong},
  \citenamefont {Zhu},\ and\ \citenamefont {You}}]{dong12}%
  \BibitemOpen
  \bibfield  {author} {\bibinfo {author} {\bibfnamefont {Y.-L.}\ \bibnamefont
  {Dong}}, \bibinfo {author} {\bibfnamefont {S.-Q.}\ \bibnamefont {Zhu}}, \
  and\ \bibinfo {author} {\bibfnamefont {W.-L.}\ \bibnamefont {You}},\ }\href
  {\doibase 10.1103/PhysRevA.85.023833} {\bibfield  {journal} {\bibinfo
  {journal} {Phys. Rev. A}\ }\textbf {\bibinfo {volume} {85}},\ \bibinfo
  {pages} {023833} (\bibinfo {year} {2012})}\BibitemShut {NoStop}%
\bibitem [{\citenamefont {Watts}\ and\ \citenamefont
  {Strogatz}(1998)}]{watts98}%
  \BibitemOpen
  \bibfield  {author} {\bibinfo {author} {\bibfnamefont {D.~J.}\ \bibnamefont
  {Watts}}\ and\ \bibinfo {author} {\bibfnamefont {S.~H.}\ \bibnamefont
  {Strogatz}},\ }\href@noop {} {\bibfield  {journal} {\bibinfo  {journal}
  {Nature (London)}\ }\textbf {\bibinfo {volume} {393}},\ \bibinfo {pages}
  {440} (\bibinfo {year} {1998})}\BibitemShut {NoStop}%
\bibitem [{\citenamefont {Newman}(2001)}]{newman01}%
  \BibitemOpen
  \bibfield  {author} {\bibinfo {author} {\bibfnamefont {M.~E.~J.}\
  \bibnamefont {Newman}},\ }\href {\doibase 10.1103/PhysRevE.64.016132}
  {\bibfield  {journal} {\bibinfo  {journal} {Phys. Rev. E}\ }\textbf {\bibinfo
  {volume} {64}},\ \bibinfo {pages} {016132} (\bibinfo {year}
  {2001})}\BibitemShut {NoStop}%
\bibitem [{\citenamefont {Latora}\ and\ \citenamefont
  {Marchiori}(2001)}]{latora01}%
  \BibitemOpen
  \bibfield  {author} {\bibinfo {author} {\bibfnamefont {V.}~\bibnamefont
  {Latora}}\ and\ \bibinfo {author} {\bibfnamefont {M.}~\bibnamefont
  {Marchiori}},\ }\href {\doibase 10.1103/PhysRevLett.87.198701} {\bibfield
  {journal} {\bibinfo  {journal} {Phys. Rev. Lett.}\ }\textbf {\bibinfo
  {volume} {87}},\ \bibinfo {pages} {198701} (\bibinfo {year}
  {2001})}\BibitemShut {NoStop}%
\bibitem [{\citenamefont {Kim}\ \emph {et~al.}(2003)\citenamefont {Kim},
  \citenamefont {Hong},\ and\ \citenamefont {Choi}}]{kim03}%
  \BibitemOpen
  \bibfield  {author} {\bibinfo {author} {\bibfnamefont {B.~J.}\ \bibnamefont
  {Kim}}, \bibinfo {author} {\bibfnamefont {H.}~\bibnamefont {Hong}}, \ and\
  \bibinfo {author} {\bibfnamefont {M.~Y.}\ \bibnamefont {Choi}},\ }\href
  {\doibase 10.1103/PhysRevB.68.014304} {\bibfield  {journal} {\bibinfo
  {journal} {Phys. Rev. B}\ }\textbf {\bibinfo {volume} {68}},\ \bibinfo
  {pages} {014304} (\bibinfo {year} {2003})}\BibitemShut {NoStop}%
\bibitem [{\citenamefont {M\"ulken}\ \emph {et~al.}(2007)\citenamefont
  {M\"ulken}, \citenamefont {Pernice},\ and\ \citenamefont
  {Blumen}}]{mulken07}%
  \BibitemOpen
  \bibfield  {author} {\bibinfo {author} {\bibfnamefont {O.}~\bibnamefont
  {M\"ulken}}, \bibinfo {author} {\bibfnamefont {V.}~\bibnamefont {Pernice}}, \
  and\ \bibinfo {author} {\bibfnamefont {A.}~\bibnamefont {Blumen}},\ }\href
  {\doibase 10.1103/PhysRevE.76.051125} {\bibfield  {journal} {\bibinfo
  {journal} {Phys. Rev. E}\ }\textbf {\bibinfo {volume} {76}},\ \bibinfo
  {pages} {051125} (\bibinfo {year} {2007})}\BibitemShut {NoStop}%
\bibitem [{\citenamefont {Andrade}\ \emph {et~al.}(2005)\citenamefont
  {Andrade}, \citenamefont {Herrmann}, \citenamefont {Andrade},\ and\
  \citenamefont {da~Silva}}]{andrade05}%
  \BibitemOpen
  \bibfield  {author} {\bibinfo {author} {\bibfnamefont {J.~S.}\ \bibnamefont
  {Andrade}}, \bibinfo {author} {\bibfnamefont {H.~J.}\ \bibnamefont
  {Herrmann}}, \bibinfo {author} {\bibfnamefont {R.~F.~S.}\ \bibnamefont
  {Andrade}}, \ and\ \bibinfo {author} {\bibfnamefont {L.~R.}\ \bibnamefont
  {da~Silva}},\ }\href {\doibase 10.1103/PhysRevLett.94.018702} {\bibfield
  {journal} {\bibinfo  {journal} {Phys. Rev. Lett.}\ }\textbf {\bibinfo
  {volume} {94}},\ \bibinfo {pages} {018702} (\bibinfo {year}
  {2005})}\BibitemShut {NoStop}%
\bibitem [{\citenamefont {de~Oliveira}\ \emph {et~al.}(2009)\citenamefont
  {de~Oliveira}, \citenamefont {de~Moura}, \citenamefont {Lyra}, \citenamefont
  {Andrade},\ and\ \citenamefont {Albuquerque}}]{oliveira09}%
  \BibitemOpen
  \bibfield  {author} {\bibinfo {author} {\bibfnamefont {I.~N.}\ \bibnamefont
  {de~Oliveira}}, \bibinfo {author} {\bibfnamefont {F.~A. B.~F.}\ \bibnamefont
  {de~Moura}}, \bibinfo {author} {\bibfnamefont {M.~L.}\ \bibnamefont {Lyra}},
  \bibinfo {author} {\bibfnamefont {J.~S.}\ \bibnamefont {Andrade}}, \ and\
  \bibinfo {author} {\bibfnamefont {E.~L.}\ \bibnamefont {Albuquerque}},\
  }\href {\doibase 10.1103/PhysRevE.79.016104} {\bibfield  {journal} {\bibinfo
  {journal} {Phys. Rev. E}\ }\textbf {\bibinfo {volume} {79}},\ \bibinfo
  {pages} {016104} (\bibinfo {year} {2009})}\BibitemShut {NoStop}%
\bibitem [{\citenamefont {Andrade}\ and\ \citenamefont
  {Herrmann}(2005)}]{andrade05mag}%
  \BibitemOpen
  \bibfield  {author} {\bibinfo {author} {\bibfnamefont {R.~F.~S.}\
  \bibnamefont {Andrade}}\ and\ \bibinfo {author} {\bibfnamefont {H.~J.}\
  \bibnamefont {Herrmann}},\ }\href {\doibase 10.1103/PhysRevE.71.056131}
  {\bibfield  {journal} {\bibinfo  {journal} {Phys. Rev. E}\ }\textbf {\bibinfo
  {volume} {71}},\ \bibinfo {pages} {056131} (\bibinfo {year}
  {2005})}\BibitemShut {NoStop}%
\bibitem [{\citenamefont {Cardoso}\ \emph {et~al.}(2008)\citenamefont
  {Cardoso}, \citenamefont {Andrade},\ and\ \citenamefont {Souza}}]{cardoso08}%
  \BibitemOpen
  \bibfield  {author} {\bibinfo {author} {\bibfnamefont {A.~L.}\ \bibnamefont
  {Cardoso}}, \bibinfo {author} {\bibfnamefont {R.~F.~S.}\ \bibnamefont
  {Andrade}}, \ and\ \bibinfo {author} {\bibfnamefont {A.~M.~C.}\ \bibnamefont
  {Souza}},\ }\href {\doibase 10.1103/PhysRevB.78.214202} {\bibfield  {journal}
  {\bibinfo  {journal} {Phys. Rev. B}\ }\textbf {\bibinfo {volume} {78}},\
  \bibinfo {pages} {214202} (\bibinfo {year} {2008})}\BibitemShut {NoStop}%
\bibitem [{\citenamefont {Souza}\ and\ \citenamefont
  {Herrmann}(2007)}]{andrem07}%
  \BibitemOpen
  \bibfield  {author} {\bibinfo {author} {\bibfnamefont {A.~M.~C.}\
  \bibnamefont {Souza}}\ and\ \bibinfo {author} {\bibfnamefont
  {H.}~\bibnamefont {Herrmann}},\ }\href {\doibase 10.1103/PhysRevB.75.054412}
  {\bibfield  {journal} {\bibinfo  {journal} {Phys. Rev. B}\ }\textbf {\bibinfo
  {volume} {75}},\ \bibinfo {pages} {054412} (\bibinfo {year}
  {2007})}\BibitemShut {NoStop}%
\bibitem [{\citenamefont {Xu}\ \emph {et~al.}(2008)\citenamefont {Xu},
  \citenamefont {Li},\ and\ \citenamefont {Liu}}]{xu08}%
  \BibitemOpen
  \bibfield  {author} {\bibinfo {author} {\bibfnamefont {X.-P.}\ \bibnamefont
  {Xu}}, \bibinfo {author} {\bibfnamefont {W.}~\bibnamefont {Li}}, \ and\
  \bibinfo {author} {\bibfnamefont {F.}~\bibnamefont {Liu}},\ }\href {\doibase
  10.1103/PhysRevE.78.052103} {\bibfield  {journal} {\bibinfo  {journal} {Phys.
  Rev. E}\ }\textbf {\bibinfo {volume} {78}},\ \bibinfo {pages} {052103}
  (\bibinfo {year} {2008})}\BibitemShut {NoStop}%
\bibitem [{\citenamefont {Boyd}(1973)}]{boyd73}%
  \BibitemOpen
  \bibfield  {author} {\bibinfo {author} {\bibfnamefont {D.~W.}\ \bibnamefont
  {Boyd}},\ }\href@noop {} {\bibfield  {journal} {\bibinfo  {journal} {Canadian
  Journal of Mathematics}\ }\textbf {\bibinfo {volume} {25}},\ \bibinfo {pages}
  {303} (\bibinfo {year} {1973})}\BibitemShut {NoStop}%
\bibitem [{Note1()}]{Note1}%
  \BibitemOpen
  \bibinfo {note} {The AN has a self-similar structure where each subnetwork,
  generated by 3 new corner nodes, locally preserves the AN recurrence but
  breaks the degree group symmetry of the original AN (see Fig. \ref
  {fig2}).}\BibitemShut {Stop}%
\end{thebibliography}
\providecommand{\noopsort}[1]{}\providecommand{\singleletter}[1]{#1}%

\end{document}